# Magnetic, Optoelectronic, and Rietveld Refined Structural Properties of Al3+ Substituted Nanocrystalline Ni-Cu Spinel Ferrites: An Experimental and DFT Based Study.

N. Hasan[a*], S. S. Nishat[b], S. Sadman[c], M. R. Shaown[d], M. A. Hoque[e], M. Arifuzzaman[f g], A. Kabir[c]

[a]Department of Electrical and Computer Engineering, North South University, Dhaka-1229, Bangladesh

[b]Materials Science and Engineering, Rensselaer Polytechnic Institute, Troy, NY 12180, United States

[c]Department of Physics, University of Dhaka, Dhaka 1000, Bangladesh

[d]Department of Industrial and Production Engineering, Bangladesh University of Textiles, Dhaka-1208, Bangladesh

[e]Bangladesh Council of Scientific and Industrial Research, Dhaka-1205, Bangladesh

[f]Department of Electronic Materials Engineering, Research School of Physics, The Australian National University, Canberra ACT 2600, Australia

[g]Department of Mathematics and Physics, North South University, Dhaka-1229, Bangladesh

[c]Department of Physics, University of Dhaka, Dhaka 1000, Bangladesh

*Corresponding author: **alamgir.kabir@du.ac.bd**, **md.arifuzzaman01@northsouth.edu**

[a]nazmul.hasan05@northsouth.edu

[b]shahriyar007@gmail.com

[c]sarker.md.sadman@gmail.com

[d]mrshaownbut44@gmail.com

[e] azizphy0043bcsir@gmail.com

[f]md.arifuzzaman01@northsouth.edu

[c]alamgir.kabir@du.ac.bd

Magnetic, Optoelectronic, and Rietveld Refined Structural Properties of Al3+ Substituted Nanocrystalline Ni-Cu Spinel Ferrites: An Experimental and Dft Based Study.

**Abstract**

The nanocrystalline $Ni_{0.7}Cu_{0.3}Al_xFe_{2-x}O_4$ (x=0.00: 0.02: 0.10) are prepared through the sol-gel auto combustion route. The structural, surface morphology, magnetic and optoelectronic properties of $Al^{3+}$ substituted Ni-Cu spinel ferrites have been reported. The crystallinity, phase structure, and structural parameters of the synthesized nanoparticles (NPs) have been determined through x-ray diffraction (XRD) and further refined by maneuvering the Rietveld refinement approach. Both XRD and Rietveld confirm the single phase cubic spinel structure of the investigated materials. Microstructural surface morphology study also confirms the formation of NPs in the highly crystalline state with a narrow size distribution. The Rietveld-refined average crystallite size of the $Al^{3+}$ doped Ni-Cu ferrite nanoparticles falls in the range (61 – 71 nm), and the average grain size is found to vary from 59 to 65 nm. All other structural parameters refined by the Rietveld refinement analysis are corroborated to single-phase cubic spinel formation of the NPs. Leveraging a vibrating sample magnetometer (VSM), the consequence of $Al^{3+}$ substitution on the magnetic parameters is studied. The saturation magnetization ($M_S$) and Bohr magneton are found to decrease with $Al^{3+}$ substitution. The Remanence ratio and coercivity ($H_C$) are observed to be very low, suggesting the materials are soft ferromagnetic. First-principle calculations were carried out using the density functional theory (DFT) to demonstrate the optoelectronic behavior of the materials. The electronic bandgap is found low as $E_g$=2.99eV for the explored materials with observing defect states at 0.62eV. The optoelectronic properties of $Al^{3+}$ substituted Ni-Cu ferrite NPs have been characterized through the DFT simulation for the first time, demonstrating their potentiality for optoelectronic device applications. The materials' optical anisotropy is observed along the x-axis, which manifests their tunability through light-matter interaction.

1. **Introduction**

Ferrites are classified as magnetic materials with astounding electromagnetic, dielectric, and other functional properties. Notably, the size and shape-dependent tunable properties of nanocrystalline ferrites (NCFs) make them promising for multi-purpose high-frequency applications. Their potential applications have been reflected through their tunable dielectric and electromagnetic properties over the last decade, which considered them suitable for a range of electronic and biomedical applications such as multi-layer chip inductors, magnetic sensors, high-density magnetic storage devices, isolators, microwave devices, wireless power transfer, hyperthermia, drug delivery, magnetic resonance imaging, gene therapy and delivery, DNA and RNA separation, and ferrofluids and so on [1–5]. As reported in [6], spinel ferrites (metal oxide semiconductors) are proven as the superior class of magnetic materials possessing the properties of high sensitivity, fast response, protracted stability, and low cost. Moreover, nanocrystalline spinel ferrite thin films have been focused in recent years on understanding the magneto-optical behaviors of annealing at comparatively low temperatures [7]. Furthermore, nanocrystalline ferrite materials are latterly being used in the production of thinner EM wave absorbers with a broader absorption bandwidth to sustain a safe and stable environment for both devices and lives by minimizing electromagnetic wave interference damage [8,9].

In $[A][B]_2[O]_4$ nano-spinel ferrites, the hetero-structure nature is highly essential for application purposes, as cation distributions over tetrahedral [A] and octahedral [B] lattice sites, as well as their inter-site exchange interactions eventually modify their magnetic characteristics [8,10]. Among various NCFs, nickel ferrites are magnetically soft materials in nature, which offer a range of practical applications due to their high magnetic permeability, moderate saturation magnetization, low eddy current loss, high resistivity, low dielectric tangent, suitable optical band gap, and high mechanical strength [9,11–13]. During the synthesis of ferrite nanoparticles, some distinct characteristics (i.e., high surface-to-volume ratio, small-size effect, and quantum tunneling effect) are observed due to bulk-to-nano-scale transition that leads to the exalted physical, magnetic, and optoelectronic properties of the materials. Consequently, ferrites' structural and magnetic properti are influenced by a range of factors such as synthesis methods, doping, processing time, sintering temperature, grain size, purity, and sintering resources [14–19].

It is noteworthy to mention that selecting a suitable substituent in forming a compatible ferrite sample is vital for fine-tuning the materials' physical properties and ameliorating the applications at a broad range [20,21]. Here, $Al^{3+}$ is chosen as the substituent since it undergoes a phase transition via a reduction in the symmetry of ferrite crystals as the effect of the Jahn-Teller effect, which in turn improves the materials' electromagnetic properties [22,23]. Moreover, $Al^{3+}$ incorporation in ferrites has a standing of improving the crystallinity with maintaining the homogeneity of magnetic nanoparticles (MNPs) [24–28]. To synthesize MNPs, the sol-gel approach is widely used among other approaches because of its improved control over powder morphology, homogeneity, and elemental composition, providing a narrow particle size distribution at relatively low temperatures. It also provides a uniformly nano-sized metal cluster, which is crucial for improving the properties of nanoparticles for high-frequency (HF) device applications [29].

Several attempts have been made sporadically to examine the structural, electrical, morphological, magnetic and dielectric characteristics of Ni-based ferrite NPs. Research in spinel ferrites is still advancing, substituting several atoms as the dopant in A and B-sites to improve their physical, dielectric, and magnetic properties. *V. A. Bharati et al.* [30] investigated the effect of $Al^{3+}$ and $Cr^{3+}$ parallel doping on the structural, morphological, and magnetic properties of Ni ferrite NPs. In [31], *K. Bashir et al.* studied electrical and dielectric properties of $Cr^{3+}$ doped Ni-Cu ferrite N', demonstrating thematerials' potential HF applications and photocatalytic activity. *Le-Zhong Li et al.* [32] studied $Al^{3+}$ substituted Ni-Zn-Co ferrites and found that saturation magnetization dropped dramatically and dc resistivity increased for $Al^{3+}$ substitution with x>0.10. The morphological and magneto-optical parameters of Ni ferrite NPs were investigated in [33], where the authors reported the bandgap, $E_g$ = 1.5 eV, and the decreasing trend was observed in the variation of saturation magnetization and Tc with $Al^{3+}$ content. *N. Jahan et al.* [34] assessed the consequences of diamagnetic aluminum ($Al^{3+}$) substitution on the morphological and magnetic properties of Ni-Zn-Co spinel ferrites fabricated using the conventional ceramic technique. They reported a steady decrease in lattice constant with increasing $Al^{3+}$ content and noted the maximum saturation magnetization ($M_s$) of 93.06 emu/g at x = 0.12. The structural, optical, magnetic, and photocatalytic behavior of $Al^{3+}$ substituted nickel-ferrites were reported in [35], which revealed n excellent photocatalytic activity showing the optical bandgap ranges between 1.60 and 1.89 eV. *Q. Khan et al.* [36] scrutinized the influence of inserting $Al^{3+}$ on the structural and dielectric behavior for Ni-Cu spinel ferrites, divulging a maximum dielectric loss of 0.4 at 2.5 GHz. Consequently, various groups explored the impact of Al doping on spinel ferrite nanoparticles' characteristics. Mn-Ni-Zn ferrite [37], Ni-Co ferrite [38], Mn–Zn ferrite [39], Co-Zn ferrites [40], Ni-Mn-Co [41], and Ni-Zn ferrite [42,43]. According to the DFT study [44], mixed spinel ferrites exhibited half-metallic properties, whereas semiconducting behavior showed pure compositions. Substitution of transitional atom content in spinel ferrite enhances the lattice parameter linearly, whereas a decrement in magnetization was observed with the weakening of the super-exchange effect in A, and B sites, as examined through DFT study in [45]. Another study employed first-principle GGA+U energy calculations for $NiFe_2O_4$ to examine the sensitivity of the cation distribution with strain modulation [46]. However, tunability in optical properties by bandgap modulation within spinel ferrite as studied through DFT calculations can offer potential for storage and photovoltaics, multifunctional materials and devices applications [47]. Moreover, DFT investigations are gradually increasing to tune the optoelectronic properties of various spinel ferrite structures, *viz.* $MgFe_2O_4$ [48], $ZnFe_2O_4$ [49], $CoFe_2O_4$ [50], $VFe_2O_4$ [51].

In [52], we performed on structural, dielectric, and electrical transport properties for $Al^{3+}$ substitution (x=0.00 to 0.10, in the step of 0.02) of nanocrystalline $Ni_{0.7}Cu_{0.3}Al_xFe_{2-x}O_4$. However, the synthesized nano spinel ferrites' Rietveld-refined structural characteristics and magnetic properties, as well as the DFT-based optoelectronic properties for such a mixed spinel ferrite structure, have not yet been reported. Therefore, this study aims to investigate how $Al^{3+}$ incorporation affects the structural (Rietveld refinement) and magnetic properties of sol-gel produced Ni-Cu ferrite NPs. The optoelectronic performances are also analyzed using the first-principle density functional theory (DFT) simulations for $Ni_{0.7}Cu_{0.3}Al_xFe_{2-x}O_4$ (x=0.06) spinel ferrite structure.

2. **Experimental and Computation Details**

2.1 **Materials preparation**

The nanocrystalline powder samples of $Ni_{0.7}Cu_{0.3}Al_xFe_{2-x}O_4$, with x varying as 0:0.02:0.1, were synthesized via the sol-gel route. Analytical grade of nickel (II) nitrate ($Ni(NO_3)_2$), copper (II) nitrate ($Cu(NO_3)_2$), iron (III) nitrate hexahydrate ($Fe(NO_3)_3.9H_2O$), and aluminum (V) nitrate $Al(NO_3)_3.9H_2O$ were used as the raw materials. These metal nitrates were dissolved in de-ionized water with adding a few drops of ethanol in a 1:2 molar ratio to obtain the initial solution with keeping its pH value at 7. The dry gel was obtained by vigorously swirling metal nitrates at 70ºC in a thermostatic water bath, then dried for 5 hours in a 200ºC electric oven. Following the process, the resultant compositions were burned and grounded before sintering at specific temperatures, and the self-ignition process progressively transformed it into fluffy-loose powder. The yielded fluffy-loose powder was annealed at 700ºC for an additional 5 hours to obtain the highly crystalline materials without impurity. The powder was homogenized further by hand-milling in a mortar to assemble disk-shaped samples. Afterward, the nanocrystalline powder was condensed into disk-like forms using a 65 MPa hydraulic press for 2 minutes. The processed samples had a diameter of 12.02 mm, and a thickness of 2.3 mm. Structure and magnetic properties were measured using the annealed powder samples.

## 2.2 Characterizations and properties measurements

Characterization of the synthesized spinel ferrite nanoparticles has been performed using a variety of techniques such as x-ray diffraction (XRD), field emission scanning electron microscopy (FESEM), energy dispersive x-ray analysis (EDX), transmission electron microscopy (TEM), vibrating sample magnetometer (VSM) and UV-Vis spectroscopy. Detail analysis for structural, electrical, and morphology using XRD, FESEM, EDX are presented in our previous study [52]. The structural properties of the prepared ferrites powder (i.e., lattice parameters (a), the crystallite size (D), displacement density, and so on) were investigated employing an x-ray diffractometer with Cu-K ($\lambda$= 1.5418 Å) radiation. Synthesized nanoparticles' magnetic properties were measured using a Vibrating Sample Magnetometer (VSM; Micro Sense, EV9). The following relationship has been used to investigate the net magnetic moment [53]:

$$\mu_B = \frac{M_1 * M_S}{5585} \quad (1)$$

where $M_1$ indicates the molecular weight of the samples, and $M_s$ denotes the magnetic saturation. Magnetic coercive force (coercivity) is a material's resistance to demagnetization since it allows us to study the material's magnetic properties than magnetization resistance, and it follows the relationship as [54]:

$$H_c = \frac{0.96 K}{M_s} \quad (2)$$

, where $K = \frac{Hc * Ms}{096}$ is used in Eq. (2) to determine the anisotropic constant.

## 2.3. Computational method

In order to learn about the mixed spinel ferrite ' 'material's light-matter interaction potentials, its optoelectronic properties were investigated with the spin-polarized Density Function Theory (DFT) [55] using VASP 6.1 (Vienna *Ab Initio* Simulation Package) [56–60]. Using the Local Spin Density Approximation (LSDA) and the Generalized Gradient Approximation (GGA) with Perdew-Burke-Ernzerhof (PBE) exchange correlation functions, the electronic charge density was optimized [61]. Pseudopotentials from the VASP library were used to describe each atom; these potentials were built on plane-wave basis sets obtained using the projector-augmented wave (PAW) approach and were parameterized for each formalism (LSDA and GGA) [62,63]. It is well-established that these pseudopotentials are more accurate for magnetic systemshan the classical ultra-soft pseudopotentials (USPPs) [60]. Taking into account on-site Coulomb interactions with the Dudarev method (LSDA+U) [63] yielded a band gap that is more precise and in good agreement with experimental data. The electronic self-consistence force convergence threshold was considered $1\times10^{-7}$ eV [64]. The Brillouin Zone was integrated using Γ-centered k-points mesh 4×4×2 generated with Monkhorst-Pack Scheme [65] and the kinetic energy cut-off was chosen 400eV. For each orbital of all atoms, partial occupancy was chosen by Gaussian smearing with a smearing width of 0.05eV to integrate the Brillouin Zone. The band structure was obtained using Wannier90 [66] interpolation to get a more accurate band structure. The number of interpolated bands was equal to the number of bands obtained from the plane wave basis code (VASP).

**3. Results and discussion**

**3.1 Structural Investigations**

*X-ray diffraction*

The x-ray diffractometer (XRD) spectra of the prepared $Al^{3+}$ substituted Ni-Cu ferrite nanoparticles are portrayed in Fig. 1, where the vital peaks are reflected from the planes of (1 1 1), (2 2 0), (3 1 1), (2 2 2), (4 0 0), (4 2 2), (5 1 1) and (4 4 0). XRD peaks confirm the single-phase spinel structure of the synthesized magnetic nanomaterials in cubic shape with no other phase pesence. The crystallite size of Ni-Cu ferrite nano samples was found to vary with $Al^{3+}$ substitutions estimated by the highest diffraction (3 1 1) plane utilizing Scherrer formula. In this study, variations in the structural parameters have been observed for 2-10% doping of $Al^{3+}$ in the lattice of Ni-Cu, which comes out with no significant alteration. The evaluated structural parameters utilizing XRD are listed in Table 1 in [52].

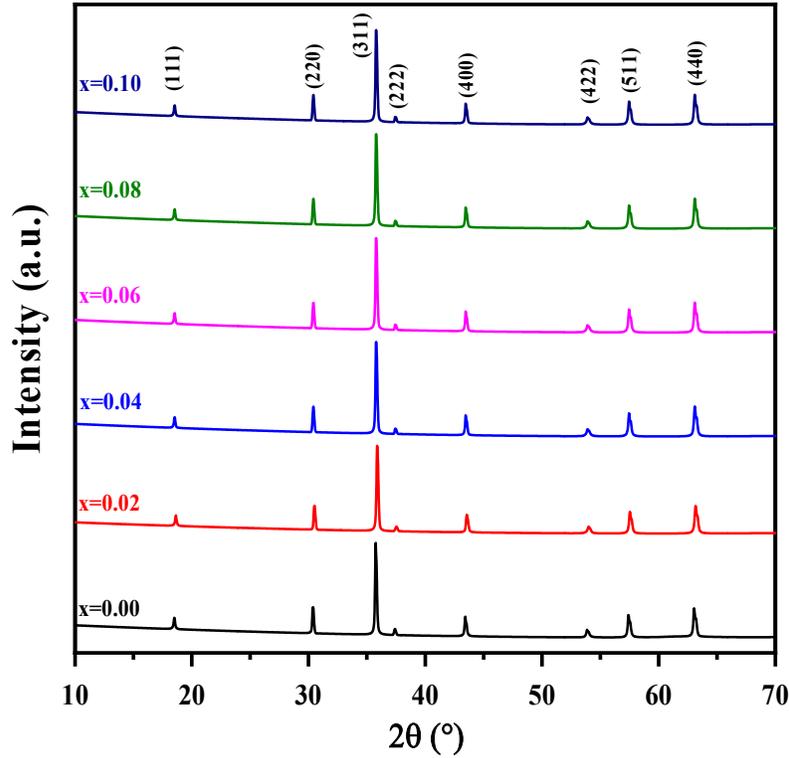

**Fig. 1.** XRD spectra profiles for the $Ni_{0.70}Cu_{0.30}Al_xFe_{2-x}O_4$ (x = 0: 0.02: 0.1) nanoparticles.

*Lattice Spacing*

Bragg's law was employed to calculate the distance between atom centers (lattice spacing) d that depends upon the direction in the lattice as following:

$$d = \frac{n\lambda}{2 S i} \qquad (3)$$

where n signifies the order of diffraction is taken as 1 as well as $\lambda$, $\theta$ is the x-ray wavelength and Bragg's angle, respectively.

*Lattice constants*

The miller indices values, i.e., (h k l) = (3 1 1), were used to calculate the lattice constants employing Eq. (4) as follows [67]:

$$a = d\sqrt{h^2 + k^2 + l^2} \quad (4)$$

Besides, the theoretical lattice constant ($a_{th}$) was estimated employing the following relation:

$$a_{th} = \frac{8}{3\sqrt{3}}\left[(r_A + R_0) + \sqrt{3}(r_B + R_0)\right] \quad (5)$$

where $R_0$ stands for the oxygen ion's radius (1.32Å) and the radius of A- and B-sites atoms denoted by $r_A$ and $r_B$ are evaluated utilizing the given formula [68]:

$$r_A = a\sqrt{3}(u - 0.25) - R_0 \quad (6)$$

$$r_B = a\left(\frac{5}{8} - u\right) - R_0 \quad (7)$$

where $u$ is the oxygen position parameter considered as 3/8 for an ideal FCC crystal. Moreover, the computed lattice parameters for each crystal-plane, displayed against the Nelson–Riley function, following:

$$F(\theta) = \frac{1}{2}\left[\frac{Co^2\theta}{Sin\theta} + \frac{Cos^2\theta}{\theta}\right] \quad (8)$$

where a diffracted line is obtained for each sample employing the diffraction angle $\theta$.

*Crystallite Size*

The average crystallite size (D) for the prepared Ni-Cu nanoparticles were estimated following Debye-Scherrer's equation from the most rising peak plane (3 1 1) [69]:

$$D = \frac{k\lambda}{\beta Cos\theta} \quad (9)$$

where the structural shape factor (Scherrer's constant) $k = 0.94$ is used for small cubic crystal, the full width at half maximum (FWHM) is denoted by $\beta$, $\lambda$ symbolizes the wavelength of incident x-rays and $\theta$ is the diffraction angle known as Bragg's angle.

*Dislocation density*

Dislocation density is the parameter used to analyze the strength and ductility of the crystal arrangement and is varied by the sample annealing. In a crystal structure, the overall dislocation length per unit volume is projected by the number of etch pits per unit area on the etched surface, as determined by equation [70]:

$$\delta = \frac{1}{D^2} \qquad (10)$$

where D represents the crystallite size. Dislocation density and particle size have followed an inverse relationship for the synthesized nanoparticles as reported in [52].

*Lattice Strain*

The unit length deforms when an object is subjected to pressure, reflecting the sample's strain. Due to the formation of defects in crystal structure and flaws in the crystal structure, the atoms' typical lattice orientations exhibit slight changes. Lattice strain quantifies the distribution of lattice constants induced by crystal defects and imperfections, such as interstitial and/or impurity atoms and lattice dislocations [71]. The following relation (Eq. 11) was used to evaluate the lattice strain for the synthesized spinel ferrites.

$$\varepsilon_{ls} = \frac{\beta}{4 \tan \theta} \qquad (11)$$

where $\beta$, and $\theta$ represent the full-width at half maximum (FWHM) of the diffraction peak and Bragg's angle, respectively.

*Micro-strain*

The most prevalent sources of deformation are dislocations, plastic deformation, point defects in the crystal structure, and abnormalities in domain boundaries, which happen in one part per million ($10^{-6}$) of the material [72]. Hence, peak broadening is a crucial aspect of micro-strain, as defined by the following equation:

$$\varepsilon_{ms} = \frac{\beta \cos \theta}{4} \qquad (12)$$

*Stacking Faults*

During crystal formation, point defects can condense into stacking faults (SF), which are distortions from the normal lattice structure induced by the layer arrangement or faults that arise in the atomic planes of the crystal [73]. The stacking fault was calculated and reported that SF values varied inversely only with the tangent of the diffraction angle $\theta$ [52].

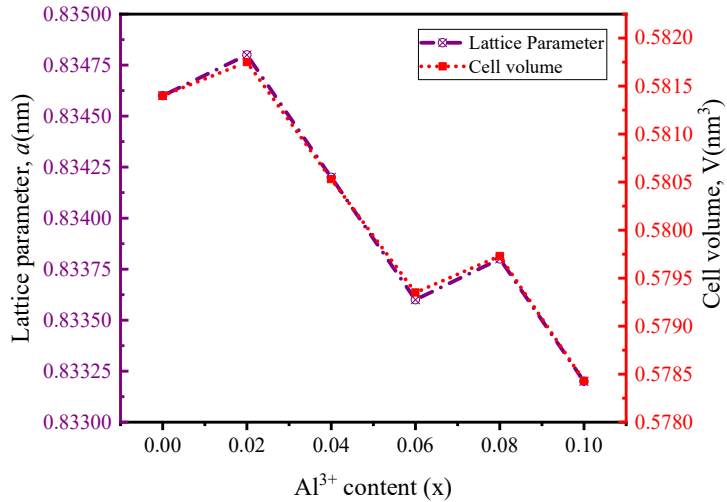

**Fig. 2:** Variation in lattice constant and cell volumes of $Ni_{0.7}Cu_{0.3}Al_xFe_{2-x}O_4$ with $Al^{3+}$ content.

Employing Eq. (9), the average crystallite size of the samples is estimated on the (h k l) = (3 1 1) plane values which are found to be in range (60-71 nm), with projecting uncertainties listed in Table 1 in [52]. The change in lattice constants and associated cell volume with increasing $Al^{3+}$ content is depicted in Fig. 2. Both the lattice parameter values and the cell volumes are observed to vary harmoniously with $Al^{3+}$ content, except for x = 0.02 and 0.08, decrease linearly with $Al^{3+}$ concentration as illustrated in Fig. 2. In variation of lattice parameters, a decreasing trend with $Al^{3+}$ content is noticed, which is justified as a larger ionic radius of $Fe^{3+}$ (0.63 Å) is replaced by $Al^{3+}$ (0.53 Å) having a smaller ionic radius following Vegard's law [74]. The literature reveals similar trends in the variation of lattice parameters [17,36,74]. The depicted downward shift of crystallite size with $Al^{3+}$ content after x=0.04 in Fig. 3 is also highlighted as high $Al^{3+}$ concentrations result in $Al^{2+}$ ions, compelling them to exchange lattice positions. This phenomenon has been appeared by a random distribution of $Al^{3+}$ and $Fe^{3+}$ ions, resulting in the increase of stress and strain in the samples. Hence, $Al^{3+}$ reduces to $Al^{2+}$ migrating from B-site to A-site after a certain amount of Al incorporation. As consequence, it is observed from Fig. 3 that the average grain sizes for the samples are increased up to x=0.04 $Al^{3+}$ incorporation and then the values were decreased.

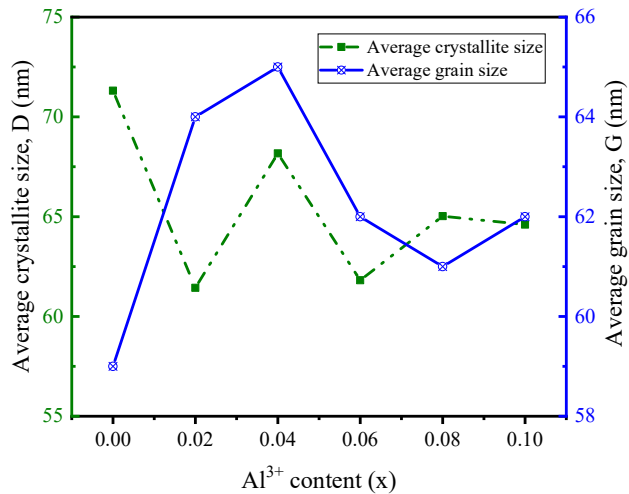

**Fig. 3:** Crystallite size and grain sizes variations with $Al^{3+}$ content.

Strain parameters help to understand the correlation between strain-induced magnetism and non-magnetic $Al^{3+}$ concentration in the ferrite nanoparticles. However, the variation in both lattice and micro strain values are portrayed in Fig. 4 to understand a clear project of the lattice distortion behavior with the $Al^{3+}$ concentrations. As depicted in Fig. 4, both of the parameters are varied in the same manner except for x=0.10, a slight splitting appears. FESEM extracted surface morphology of the investigated materials exhibited that the grains were mostly about spherical in shape and are distributed uniformly and evenly due to the separating grain boundaries as described in [52]. FESEM micrographs, on either hand, revealed multi-grain phenomena composed of grains and grain boundaries, with some agglomerations appearing due to the dipole-dipole interaction within magnetic nanoparticles and the high surface to volume ratio. The mean particle size of the samples was found to be in the nano-size range (59 - 65 nm). The intensity peaks in the Energy-dispersive X-ray spectroscopy (EDX) spectra are exacerbated by the energy gap between two electronic states generated by the cannonade of composites by the electron beams of the SEM. As reported in [52], there are no impurity peaks, reconfirming the materials' single-phase structure observed after ascertaining the proper compositional proportions of elements contained in the synthesized nanocrystalline ferrite samples.

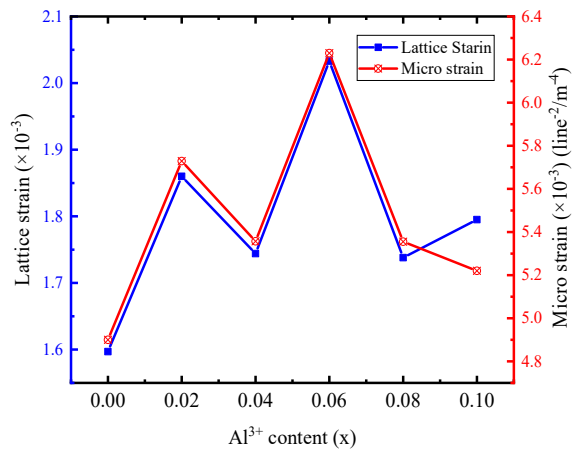

**Fig. 4:** Variation in lattice strain and micro strain of nanocrystalline $Ni_{0.7}Cu_{0.3}Al_xFe_{2-x}O_4$.

*Rietveld refinement*

The refinement of XRD extracted data is carried out through the Rietveld refinement analysis using the FullProf software as illustrated in Fig. 5. In this refining process, a nonlinear least-squares fitting method with a Pseudo-Voigt profile was utilized. Throughout the fitting procedure, instrumental and background parameters were carefully considered to refine the structural parameters and this process was repeated until a minimal residual (difference between calculated and observed intensities) was found along with achieving a good value of fitting parameter, $\chi^2$.

The Rietveld fitting parameters are presented in Table 1, with R-factors and $\chi^2$ values. For each sample's fitting, Fig. 5 displays the observed intensity ($Y_{obs}$), the estimated intensity ($Y_{cal}$), and the residual, which are obtained from the refinement process. Also, Table 1 displays the refined crystal parameters (i.e., lattice parameter, unit cell volume, and average crystallite size).

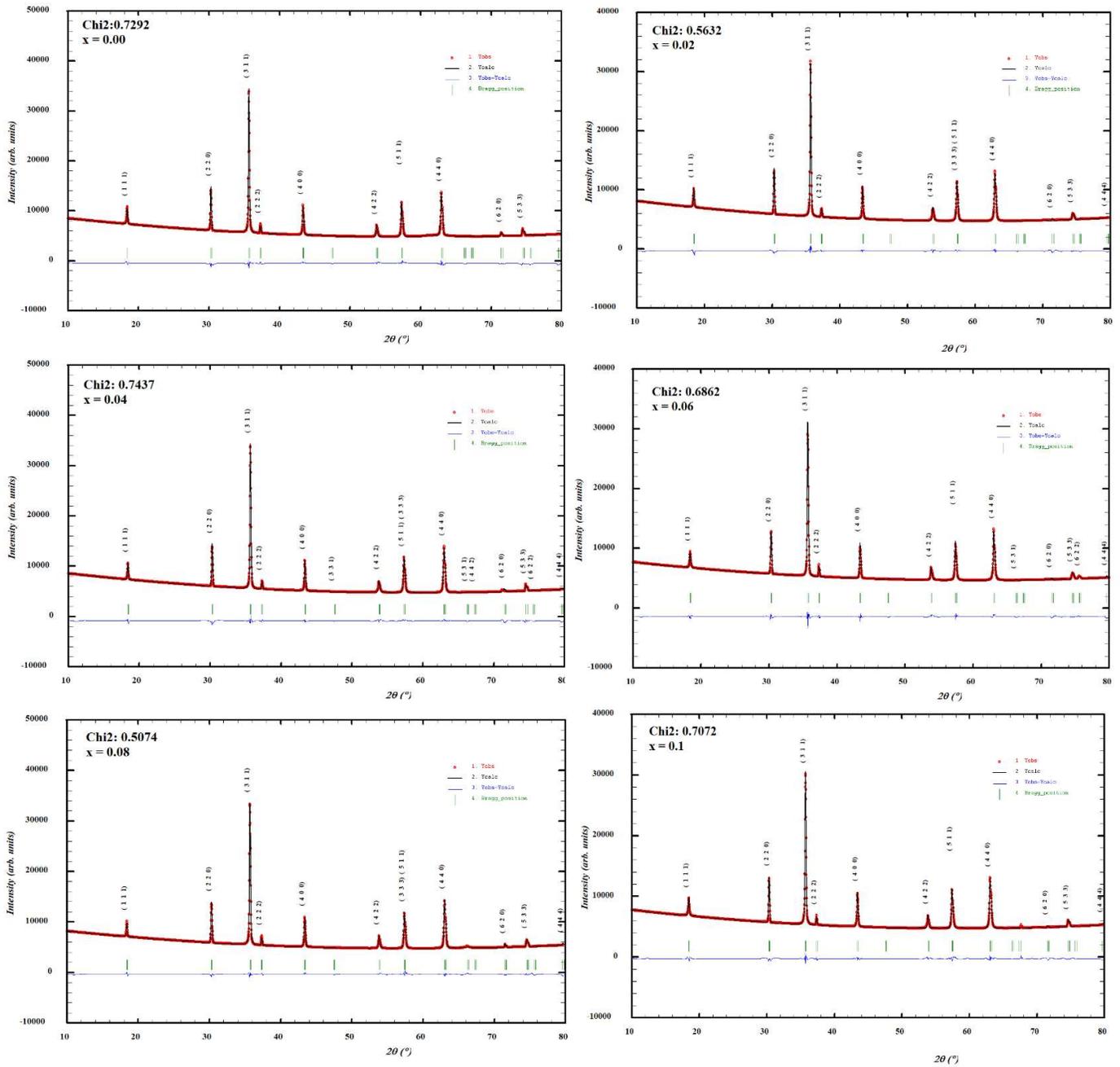

**Fig. 5:** Rietveld refinement of various samples in nanocrystalline $Ni_{0.7}Cu_{0.3}Al_xFe_{2-x}O_4$ annealed at 700 ºC.

The overlapping between the calculated and observed patterns indicate that well-fitted patterns, found in all cases in these analyses of Fig. 5. Similar to the correlation profiles, the difference between the calculated and observed patterns is almost linear with just slight fluctuations, denoting that the data is optimally-fitted. For the $Al^{3+}$ concentration in the nano-spinel ferrite compounds at different values of x (0.00, 0.02, 0.04, 0.06, 0.08 and 0.10), the goodness-of-fit values for $\chi^2$ observed are 0.7292, 0.5632, 0.7437, 0.6862, 0.5074, and 0.7072. The value of $\chi^2 = 2.00$ is considered as a better fitting indicator in the Rietveld refinement analysis [72] and thus the observed values of $\chi^2$ have accorded well with the observed fitting patterns.

Table 1. Rietveld refined fitting parameters, lattice parameters, cell volume, and average crystallite size for the $Al^{3+}$ doped Ni-Cu spinel nanoferrites.

| $Al^{3+}$ content (x) | $R_p$ | $R_{wp}$ | $R_{exp}$ | $\chi^2$ | DW-stat | $a_{exp}$(Å) | $V$(Å)$^3$ | Average Crystallite Size D (nm) |
|---|---|---|---|---|---|---|---|---|
| 0.00 | 12.5 | 8.05 | 9.43 | 0.73 | 0.166 | 8.346 | 581.40 | 71.32 |
| 0.02 | 12.1 | 7.04 | 9.38 | 0.56 | 0.108 | 8.347 | 581.75 | 61.43 |
| 0.04 | 13.4 | 7.89 | 9.15 | 0.74 | 0.125 | 8.342 | 580.53 | 68.17 |
| 0.06 | 12.7 | 7.81 | 9.43 | 0.67 | 0.295 | 8.336 | 579.35 | 61.81 |
| 0.08 | 11.6 | 6.52 | 9.15 | 0.51 | 0.154 | 8.338 | 579.73 | 65.03 |
| 0.10 | 15.5 | 8.05 | 9.56 | 0.71 | 0.175 | 8.332 | 578.43 | 64.59 |

*Hopping Lengths*

The hopping length refers to the average distance traveled by an ion from one neighboring lattice-site to another. We used the following relations to determine the hopping distance between A-sites, B-sites, and shared sites, respectively [75]:

$$L_{A-A} = \frac{a_o\sqrt{3}}{4} \quad (17)$$

$$L_{B-B} = \frac{a_o}{2\sqrt{2}} \quad (18)$$

$$L_{A-B} = \frac{a_o\sqrt{11}}{8} \quad (19)$$

To study the hopping mechanism and further cation distributions over the lattice sites, the hopping lengths for both tetrahedral and octahedral sites are evaluated following Eq. (17-19). The estimated hopping lengths for A-sites ($L_A$), B-sites ($L_B$), and shared A-B sites ($L_{A-B}$) are tabulated in Table 2 The variations in both $L_A$ and $L_B$ with $Al^{3+}$ substitution follow a similar trend as shown in Fig. 6.

As the grain size shifts, the distance between magnetic ions shifts, which in turn affects the $L_A$ and $L_B$ values [76]. Fig. 6 clearly reveals that $L_A>L_B$, suggesting that the probability of electron hopping between ions in tetrahedral A and octahedral B sites is lower than that between octahedral B-B sites. Moreover, hoping lengths also gives the insight to understand the electrical conduction characteristics since the effectiveness of the scattering process of hopping electrons directly affects the conductivity [77]. The other structural parameters as listed in Table 2- bond lengths ($d_{A\times}$ and $d_{B\times}$), tetrahedral edge ($d_{A×E}$), shared ($d_{B×E}$) and unshared ($d_{B×Eu}$) octahedral edges for the synthesized $Ni_{0.7}Cu_{0.3}Al_xFe_{2-x}O_4$ samples are evaluated using lattice parameter value $a$ and the oxygen positional parameter $u = 0.381$ Å utilizing the following equations [78]:

$$d_{A×} = a\sqrt{3}\left(u - \frac{1}{4}\right) \quad (20)$$

$$d_{B×} = a\sqrt{3u^2 - \frac{11}{4}u + \frac{43}{64}} \quad (21)$$

$$d_{A×E} = a\sqrt{2}\left(2u - \frac{1}{2}\right) \quad (22)$$

$$d_{B×E} = a\sqrt{2}(1 - 2u) \quad (23)$$

$$d_{B×Eu} = a\sqrt{4u^2 - 3u + \frac{11}{16}} \quad (24)$$

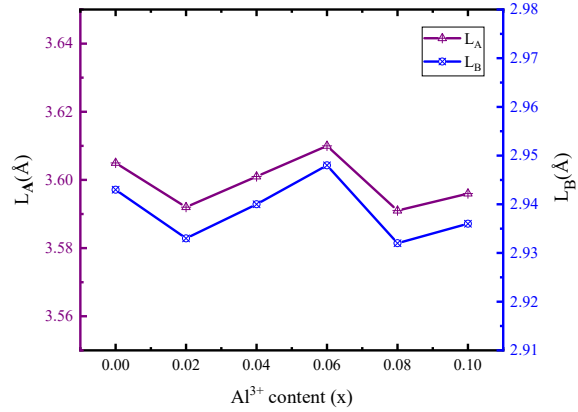

**Fig. 6:** Variation in hoping lengths ($L_A$ and $L_B$) with $Al^{3+}$ content in nanocrystalline $Ni_{0.7}Cu_{0.3}Al_xFe_{2-x}O_4$.

**Table 2.** Hopping lengths ($L_A$, $L_B$, and $L_{A-B}$), tetrahedral and octahedral bond lengths ($d_{A×}$ and $d_{B×}$), tetrahedral edge ($d_{A×E}$), shared ($d_{B×E}$) and unshared ($d_{B×EU}$) octahedral edge for the synthesized Ni-Cu ferrite nanoparticles

| $Al^{3+}$ content (x) | $L_A$(Å) | $L_B$(Å) | $L_{A-B}$(Å) | $d_{A×}$(Å) | $d_{B×}$(Å) | $d_{A×E}$(Å) | $d_{B×E}$(Å) | $d_{B×EU}$(Å) |
|---|---|---|---|---|---|---|---|---|
| 0.00 | 3.605 | 2.943 | 3.451 | 1.894 | 2.038 | 3.787 | 2.809 | 2.953 |
| 0.02 | 3.592 | 2.933 | 3.440 | 1.894 | 2.038 | 3.093 | 2.810 | 2.953 |
| 0.04 | 3.601 | 2.940 | 3.448 | 1.893 | 2.037 | 3.091 | 2.808 | 2.951 |
| 0.06 | 3.610 | 2.948 | 3.456 | 1.891 | 2.035 | 3.089 | 2.806 | 2.949 |
| 0.08 | 3.591 | 2.932 | 3.438 | 1.892 | 2.036 | 3.089 | 2.807 | 2.950 |
| 0.10 | 3.596 | 2.936 | 3.443 | 1.890 | 2.034 | 3.087 | 2.806 | 2.947 |

Transmission electron microscopy (TEM) is used to validate the structural morphology of nanocrystalline nature of the synthesized ferrites. TEM micrographs of $Ni_{0.7}Cu_{0.3}Fe_2O_4$ ferrite nanoparticles annealed at 700ºC are shown with different magnifications in Fig. 7 (A & B), where the uniformity and nano-spherical shape of the particles is evident. When the plane of the atoms is in the same direction and linear, high crystallinity is achieved, and the average d-spacing is 0.458 nm which are exemplifies perfectly through the HTEM image in Fig. 7(C). In Fig. 7(D), the selected area electron diffraction (SAED) pattern is shown to be well matched with the x-ray diffractogram results. The SAED pattern confirms that the spotty diffraction rings correspond to planes (1 1 1), (2 2 0), (3 1 1), (2 2 2), (4 0 0), (4 2 2), (5 1 1), and (4 4 0), reiterating the single spinel phase of the synthesized nanocrystalline ferrites.

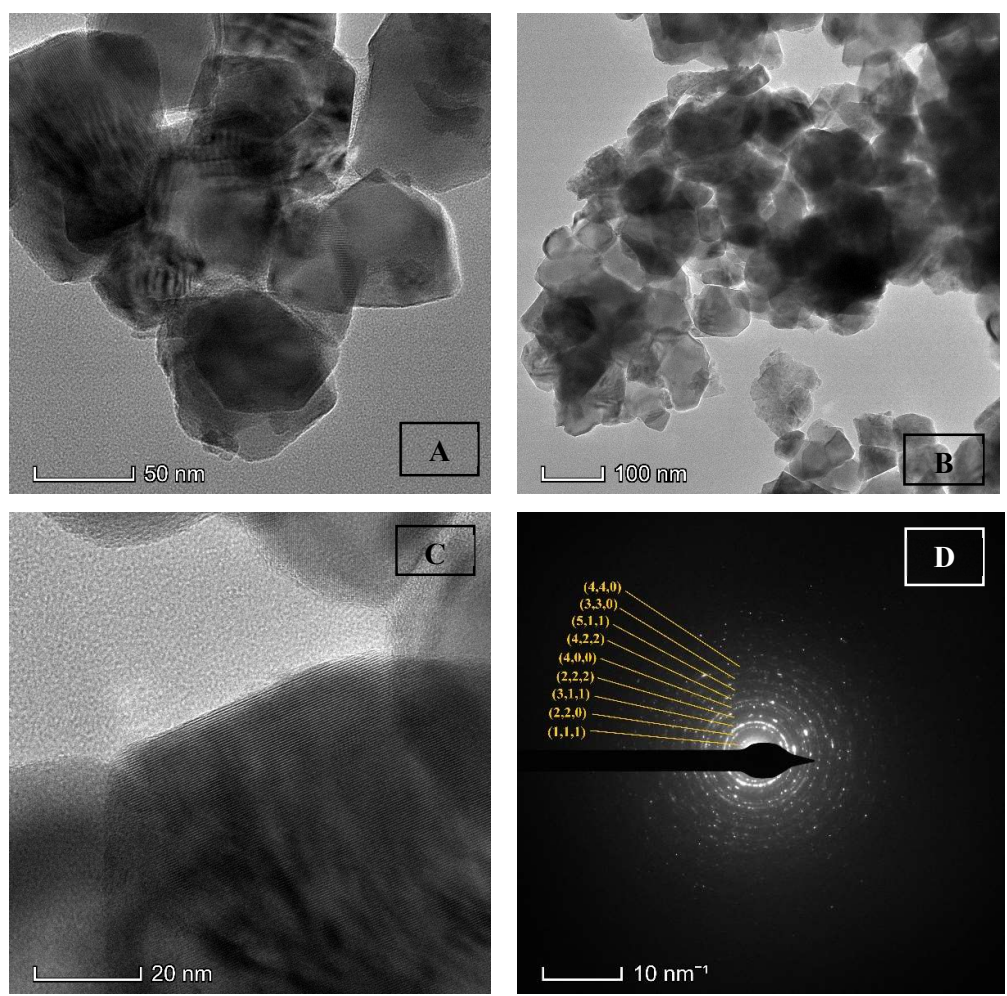

**Fig. 7:** TEM [(A) & (B)], HTEM (C) morphological micrographs, and (D) SAED patterns of nanocrystalline $Ni_{0.70}Cu_{0.30}Al_xFe_{2-x}O_4$.

### 3.2 Magnetic Properties

The magnetic properties of ferrites are affected by the composition of metal ions and their distribution in the spinel lattice. The variation in cation distribution over tetrahedral (A-site) and octahedral (B-site) lattice sites causes the variation in magnetic properties. At room temperature, the magnetic parameters of $Al^{3+}$ incorporated Ni-Cu ferrite nanoparticles are determined using the VSM method with varying applied magnetic fields (H). The M-H curves of the synthesized materials are represented in Fig. 9. From the VSM loops, the magnetic parameters (i.e., Ms, Mr, Hc, K, and so on) are calculated and they are listed in Table 4.

*M-H Hysteresis loop*

Fig. 8 depicts the magnetization variation of the investigated ferrite nanoparticles with reference to the applied magnetic field (H) (known as the M-H loop) at room temperature, which is one of the crucial parameters in determining the plausible application of magnetic materials. In Fig. 8, it is observed that magnetization (M) for all the compositions of the synthesized ferrites increases up to 0.5T, thereafter it becomes steady to rise and then appears with the saturation magnetization value for the applied field to 2T. The magnetization variation trend indicates the soft ferrimagnetic nature of the investigated Ni-Cu spinel ferrite nanoparticles. The magnetization curve was extrapolated to $\mu_0 H = 0$ to evaluate the $M_s$ for all the studied compositions. However, from the curves extracted from VSM, estimated values of magnetic parameters as- $M_s$ : saturation magnetization, $M_r$ : remanence magnetization, $H_c$ : coercivity, $M_r/M_s$ : remanence ratio, $\eta_{exp}(\mu_B)$ : experimental magnetic moment, and K : magnetic anisotropy constant are listed in Table 3.

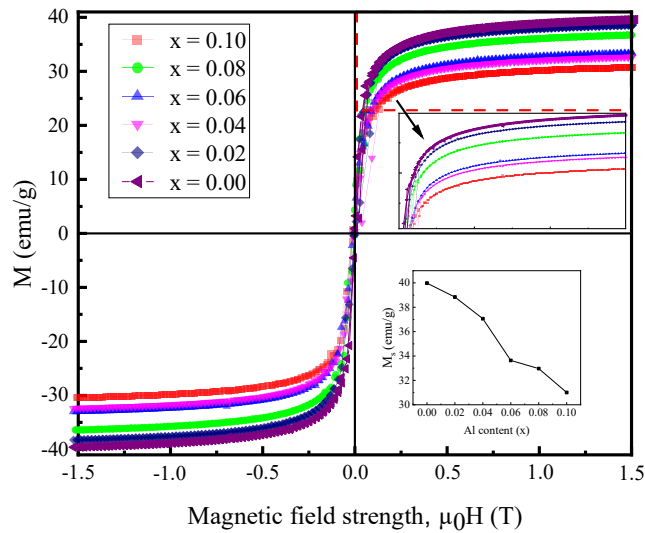

**Fig. 8:** Variation in magnetization with applied field strength of nanocrystalline $Ni_{0.7}Cu_{0.3}Al_xFe_{2-x}O_4$ at room temperature.

Fig. 9(A) demonstrates that $M_s$ steadily decreased with the gradual increase of $Al^{3+}$ substitution in composition. In the examined Ni-Cu ferrites, the cations are distributed over A- and B-sites according to Neel's two sub lattice model, where the magnetic moments of various cations are $Ni^{2+}$ (2.3 µB), $Cu^{2+}$ (1 µB), $Al^{3+}$ (0 µB), and $Fe^{3+}$ (5 µB), respectively. The insertion of $Al^{3+}$ ions might well have occupied the B sublattice, and the probable A sub lattice occupying states were very small. As a result, the number of magnetic moments on the B-site was expected to decrease as the $Al^{3+}$ concentration in the samples changed. The total magnetization, on the other hand, is equal to the difference between B-site and A-site magnetization. As a result, increasing the content of $Al^{3+}$ ions in the compositions reduces the value of $M_s$. However, as B-site is occupied by a nonmagnetic $Al^{3+}$ leads to no change in magnetization. The earlier studies [15,17] suggest that with further $Al^{3+}$ incorporation in the ferrites, exchange interaction by the crystal symmetry reduction in lattice occurs, therefore, $Cu^{2+}$ partially migrates its site from A to B, $Fe^{3+}$ shifts from B to A and $Al^{3+}$ alleviates to $Al^{2+}$ thus relocating to site A in the manner can be described as $Cu^{2+} \leftrightarrow Cu^{3+}+e-$, $Fe^{3+} +e- \leftrightarrow Fe^{2+}$ and $Al^{3+} + e- \leftrightarrow Al^{2+}$. Facts of cation distribution according to the Neel's lattice model and the exchange interactions among lattice sites ($J_{AA}$, $J_{AB}$, and $J_{BB}$) where each is influenced by oxygen ions enable it to be comprehended the trend seen in Fig 9(A) on the variation of magnetic saturation values and magnetic moment with $Al^{3+}$ concentrations [79].

Furthermore, one unpaired electron contains in $Al^{3+}$ with a magnetic moment of $\sqrt{3}$, whereas $Al^{2+}$ has no valence electrons that are unpaired. However, Oxygen seems to have two unpaired electrons in its valence state which compensate for neutralizing the charge of the samples. Hence, the lattice constant changes influencing the $J_{AB}$ interaction leads to altering the magnetic performance due to cationic redistribution between A and B sites and lattice distortion in the crystal symmetry of the ferrites [80,81]. The saturation magnetization ($M_S = M_B - M_A$) and consequently the experimental magnetic moment ($\eta_{exp}$) are found to decrease with the increase of $Al^{3+}$ content as depicted in Fig. 9(A) for destabilization of the A-B inter-site interaction. However, both parameters are shown maximum for the pristine sample with no $Al^{3+}$ content. Similar variations in such magnetic behaviors for the ferrite nanoparticles have been reported in the literature [17,28,82].

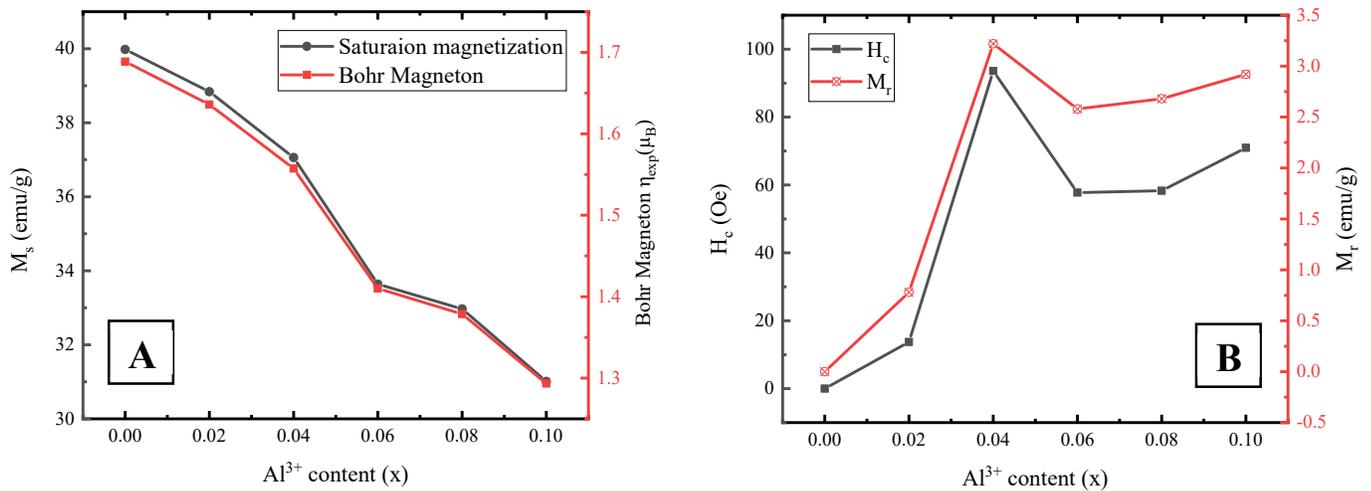

**Fig. 9:** Variations in (A) saturation magnetization and magnetic moment, (B) remanence magnetization and coercivity with $Al^{3+}$ content.

Utilizing the M-H loops evaluated values of $M_r$ and $H_C$ and listed in Table 3 as well as the variation of both parameters is presented in Fig. 9(B). As shown in Fig. 9(B), both $M_r$ and $H_C$ demonstrate a similar behavior with the incorporation of $Al^{3+}$ content. The resultant low coercivity values for the examined ferrite nanomaterials classify them as soft magnetic materials. Fig. 10 portrayed the projection of changes between the $M_r/M_s$ (remanence ratio) and K (magnetic anisotropy constant) for $Ni_{0.7}Cu_{0.3}Al_xFe_{2-x}O_4$ samples with $Al^{3+}$ substitution changes. Several factors contribute to the variation of $H_C$ and K values with $Al^{3+}$ concentration, in which magnetic domain walls and corresponding magnetic moments are significant [53]. The K values are found to be increased with $Al^{3+}$ concentration, which is attributed to the site exchange interactions among magnetic nanoparticles. The remanence ratio of the synthesized ferrites is found to be very low in a range of (0.000 – 0.094) [Table 3], indicating the presence of magnetic nanoparticles with a multi-domain nature in the samples as also reported earlier [53,83,84].

Table 3. Measured magnetic parameters for the synthesized $Ni_{0.7}Cu_{0.3}Fe_{2-x}Al_xO_4$ samples.

| $Al^{3+}$ content (x) | Ms (emu/g) | Mr (emu/g) | Hc (Oe) | Mr/Ms | $\eta_{exp}(\mu_B)$ | K (erg/Oe) |
|---|---|---|---|---|---|---|
| **0.00** | 39.98 | 0.00 | 0.00 | 0.000 | 1.6887 | 0.00 |
| **0.02** | 38.84 | 0.78 | 13.74 | 0.020 | 1.6361 | 555.90 |
| **0.04** | 37.06 | 3.22 | 93.57 | 0.087 | 1.5573 | 3612.19 |
| **0.06** | 33.64 | 2.58 | 57.76 | 0.077 | 1.4101 | 2024.01 |
| **0.08** | 32.97 | 2.68 | 58.31 | 0.081 | 1.3786 | 2002.58 |
| **0.10** | 31.01 | 2.92 | 70.94 | 0.094 | 1.2934 | 2291.51 |

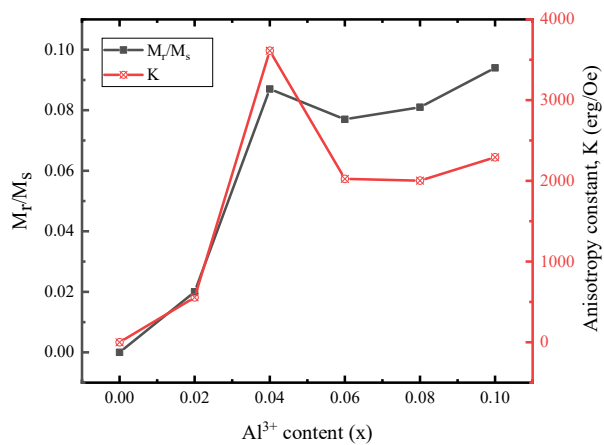

**Fig. 10:** Changes in the remanence ratio and the magnetic anisotropy constant at room temperature with varying $Al^{3+}$ concentration.

*UV-Vis Analysis:*

Optical absorption spectrum (UV–Vis) study is one of the suitable methods for understanding the optical property of the materials. Fig. 11 shows the UV-vis absorption spectra of the prepared Ni-Cu nano spinel ferrites that exhibit absorption spanned in a wide wavelength range of 200-800 nm. It is seen from Fig. 11(A) that there is a tendency to show different slopes at different wave lengths, which can be ascribed to the electron transitions between the oxygen ions and cations. To estimate the direct bandgap energies (Eg), the UV–Vis spectra were plotted; $(\alpha h\nu)^2$ vs photon energy ($E=h\nu$) using the formula: $E_{bg}$ (eV) $\leq$ 1240/$\lambda$ [85,86], and represented in Fig. 11(B). Optical absorption spectra of $Al^{3+}$ substituted $Ni_{0.7}Cu_{0.3}Fe_2O_4$ nanoparticles resulted the direct band gap energy ($E_{bg}$) 2.60 eV, 3.00 eV, 3.24 eV, 3.07 eV, 3.37 eV, and 2.80 eV respectively for x=0.00,0.02,0.04,0.06,0.08 and 0.10 concentration. It can also be seen that the UV–Vis absorption spectra exhibit a slight spline shape indicating to the localized (by means of structural defects) electronic levels above the valence band gap. Additionally, it has been found that in wide bandgap semiconductors, shallow defect levels near the conduction band or valence band do not play an effective role as a recombination site. However, the deepest level within the forbidden band is where the most effective recombination takes place [87]. It appears that the defects at the octahedral location with Al and Fe in the prepared Ni-Cu spinel ferrites could kick off recombination. However, the optical properties since associated with the crystal formation, hence lattice and micro strain along with the hopping lengths effectively play role to alter the properties.

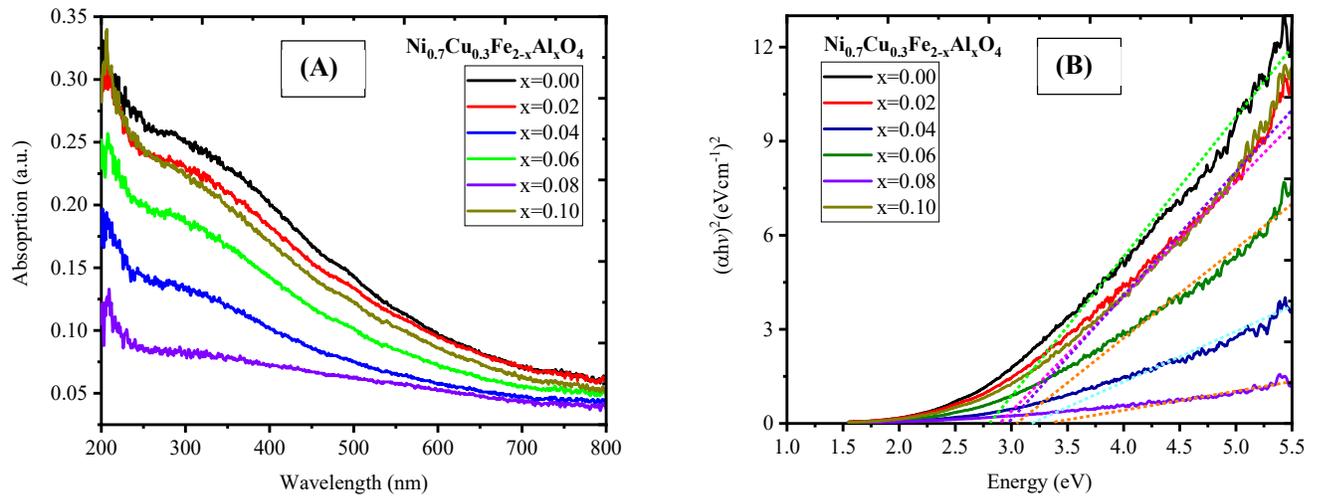

**Fig. 11:** Variations in (A) Absorptions and (B) Band gap values for Ni-Cu nano spinel ferrites with varying $Al^{3+}$ content.

**Density Functional Theory analysis**

To mimic the $Ni_{0.7}Cu_{0.3}Fe_{1.94}Al_{0.06}O_4$ sample, standard DFT simulation was performed, and the electronic and optical properties were investigated. A $2 \times 1 \times 1$ supercell was constructed from the conventional $Ni_{0.7}Cu_{0.3}Fe_2O_4$ unit cell of $Fd\bar{3}m$ space group in the Cubic crystal system. Since previous experiments reported Fe, Ni and Cu with radii 0.77Å, 0.69Å, 0.71Å respectively [88][89]. The $Al^{3+}$ ions with ionic radius ~0.48Å [88], the Cu atoms occupied the tetrahedral sites. They occupied the positions of Ni atoms, and our experiment suggests the Al doping into the structure there it occupied the tetrahedral site of Fe atoms. Therefore, five Ni atoms were then replaced with five Cu atoms to design a 30% Cu doping in the Ni occupied A-site, after that two Fe atoms were replaced with two Al atoms for the sake of Al (x=0.06) doping in the B-site and the DFT simulations were performed. Spin polarized calculations with GGA-PBE functionals were performed to relaxed the structure. Table 4 shows the structural parameters from DFT calculations. It is noticed that the lattice parameter ($a$), cell volume (V), and inter bonds are well consistent with the experimental data.

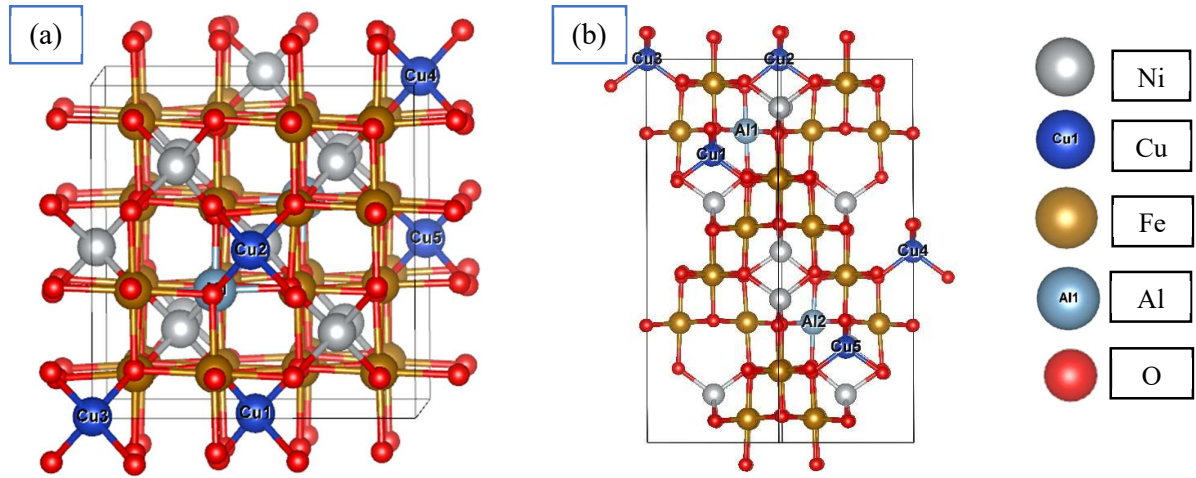

**Fig. 12:** Optimized atomic structure of the $Ni_{0.7}Cu_{0.3}Fe_{1.94}Al_{0.06}O_4$ spinel ferrite nanoparticle (a) 001 plane, (b) 111 plane.

**Table 4. DFT extracted structural optimized parameters for $Ni_{0.7}Cu_{0.3}Fe_{1.94}Al_{0.06}O_4$.**

| $Ni_{0.7}Cu_{0.3}Fe_{1.94}Al_{0.06}O_4$ | Lattice Parameter $a$ (Å) | Cell Volume V (Å³) | $L_{A-B}$ (Å) | $d_A$ (Å) | $d_B$ (Å) |
|---|---|---|---|---|---|
| | alpha 8.29 | 569.72 | Ni-Fe: 3.38<br><br>Ni-Al: 3.51<br><br>Cu-Fe: 3.55<br><br>Cu-Al: 3.51 | Ni-O: 1.96<br><br>Cu-O: 2.02 | Fe-O: 2.043<br><br>Al-O: 1.928 |

**Electronic Properties**

As we know, the standard DFT packages underestimate the band gap [90], so we used the DFT+U formalism in the Dudarev approach. The $U_{eff}$ was applied to the 3d orbitals of Fe, Ni and Cu atoms by the amount 4eV, 6.4eV and 4eV correspondingly [91] and found the electronic band gap was consistent with the experimentally observed value. The equilibrium structure obtained from the calculation was Triclinic in phase, and the lattice parameters were a=b=8.288Å, and c=17.078Å does have a good agreement with the experimentally observed data. As referred, the material to be a magnetic structure, the magnetic nature of Fe, Ni, O, Cu, and Al was introduced, and found the equilibrium magnetic moments of corresponding elements as $4.257\mu_B$, $-1.796\mu_B$, $0.259\mu_B$ and $-0.608\mu_B$. However, as the material is magnetic in nature, so the spin polarized calculations were performed. Hence the Spin-UP bands (represented in Fig. 13(a)) and Spin-Down (represented in Fig. 13(b)) bands were obtained. The Wannier90 interpolated band structure is presented in Fig. 13. From the figure, it is clearly evident that the doped material has a direct band gap at the Γ-point in both cases.

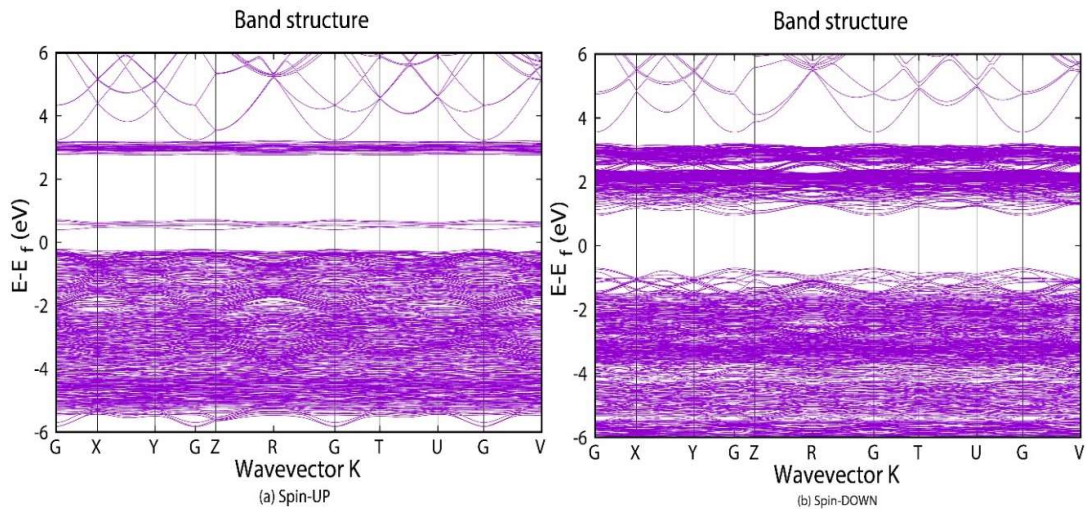

**Fig. 13**: Electronic Band Structure of $Ni_{0.7}Cu_{0.3}Fe_{1.94}Al_{0.06}O_4$ nanocrystalline ferrite.

Therefore, there can be a direct transition of electrons from the valance band to the conduction band without phonon interaction. The Spin-UP case has a band gap of 2.99eV, and the Spin-Down bands have a band gap of 1.66eV. The DFT calculated optical band gap nature and value is found in tune with the experimental data as presented in Fig. 11. The direct band gap makes the material a potential candidate for optoelectronic device applications. However, a defect band state evident in the Spin-UP band structure in the conduction band region. That defect states located 0.62eV above the valence band maxima.

To investigate the defect region, we performed the Density of States (DOS) and Partial Density of States (PDOS) calculations presented in the Fig. 14. The DOS shows that the material has a defect state near fermi region in the conduction band. This occurred due to the strong hybridization between Cu-3d orbital and O-2p orbital; a small contribution exists due to the Fe-3d orbital but no significant contribution of Al-3(s,p) orbitals can exist be observed in the defect region. Therefore, it can be concluded that this material has an opportunity to perform as a photocatalyst in color degradation. Moreover, the above experiment reports that the material is magnetic in nature, which also can be observed in the non-symmetric density of states represented in Fig. 14. From the PDOS it can be easily inferred that the magnetism is mostly driven by the Fe-3d orbital electrons and the 3d orbitals of Ni and Cu simultaneously.

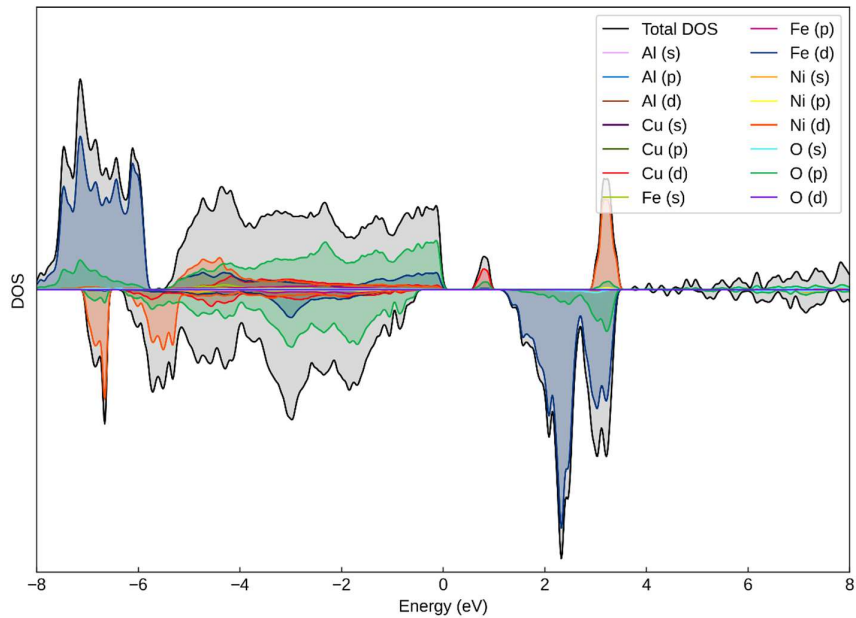

**Fig. 14**: Density of States of $Ni_{0.7}Cu_{0.3}Fe_{1.94}Al_{0.06}O_4$ nanoparticle.

**Optical Properties**

As previously mentioned, the above material has a potential optoelectronic device application. Therefore, the optical profile is crucial for this purpose. So, we did an investigation in its optical properties with the GGA-PBE+U formalism. Where the dielectric function plays the critical role, and absorption coefficient $\alpha$, reflectivity R, dielectric energy loss function L, refractive index $\eta$, optical conductivity $\sigma$ like information concealed. To perceive these optical properties of this material at-first we extracted the dielectric function and its corresponding real and imaginary parts from the Kramer-Kronig relation - $\epsilon(\omega) = \epsilon_{real}(\omega) + i\epsilon_{imag}(\omega)$ [92]. Hence, the light-matter interaction opens tremendous opportunities in case of magnetic matter, we considered polarization along three orthogonal axes-x,y,z which leads us to the optical anisotropy of the concerned material along the x-direction. The optical anisotropy of the doped material constituted in Fig. 15(a, b), which evokes the anisotropy along x-axis. In the static limit, $\omega \to 0$, $\varepsilon_{real\_x}(0) = 6.05$eV when $\varepsilon_{real\_y} = \varepsilon_{real\_z} = 5.84$eV, where $\varepsilon_{real\_i}$ is the real part of dielectric function along $i(x,y,z)$-direction. But, the imaginary part of the dielectric function at the static limit vanishes in all three cases i.e. $\varepsilon_{imag\_x} = \varepsilon_{imag\_y} = \varepsilon_{imag\_z} = 0$ eV. Therefore, the average of $\varepsilon_{real}$ is 5.91eV and the maximum of $\varepsilon_{real}$ lies at 2.86eV along the x-direction polarization and 2.92eV for polarization along y,z-directions. The absorption spectra presented in the Fig. 15(c) infers that there is no absorption below the band gap energy nor any band transition appeared.

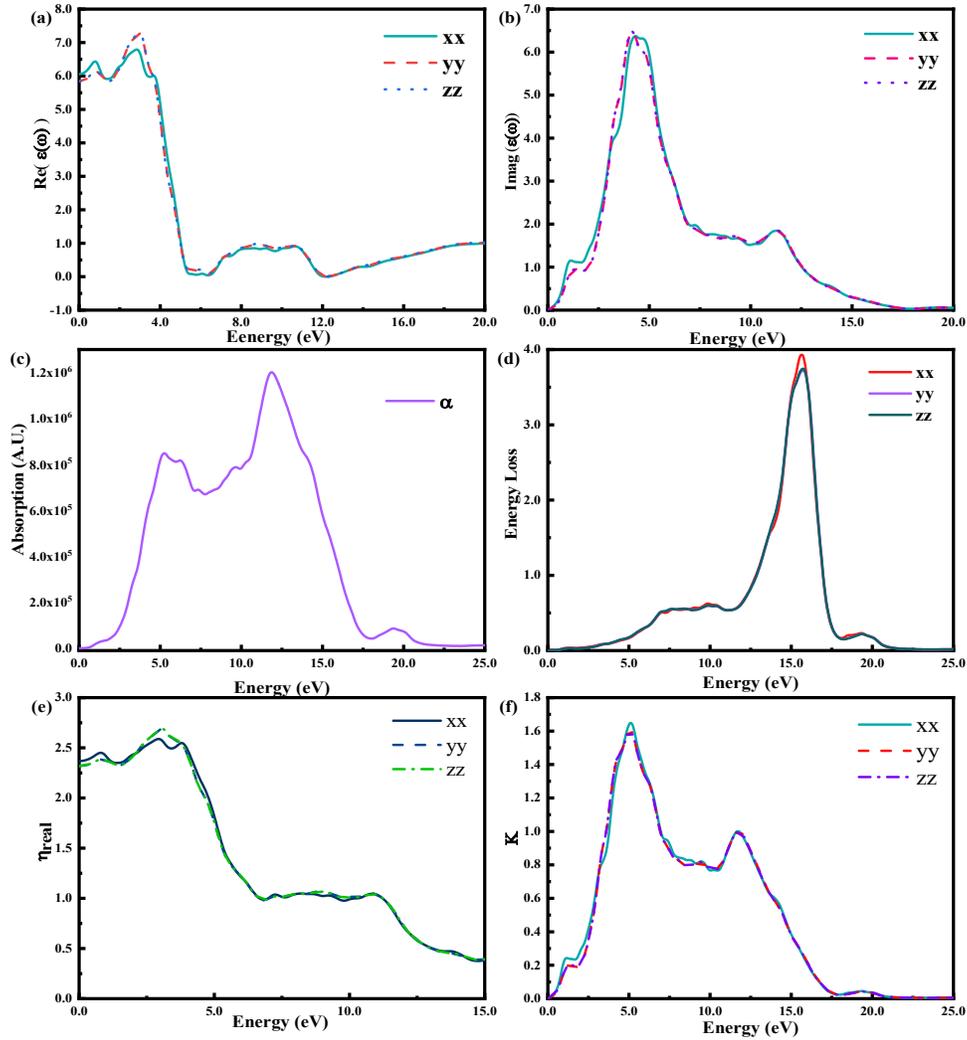

**Fig. 15**: DFT extracted (a) Real part of Dielectric Function, (b) Imaginary part of Dielectric Function, (c) Absorption spectra, (d) Dielectric Energy Loss, (e) Real pat of Refractive Index, (f) Extinction coefficient for $Ni_{0.7}Cu_{0.3}Fe_{1.94}Al_{0.06}O_4$ nanoparticle.

The dielectric energy loss plotted in Fig. 15(d), showing results that there is a sharp gain starting around 11.63eV which diminishes around 21.35eV. The sharp peak around 15.6-16.0eV corresponds to the plasmon frequency. Refractive Index of the $Al^{3+}$ (6%) doped $Ni_{0.7}Cu_{0.3}Fe_2O_4$ has been studied along three orthogonal polarization directions $x, y, z$ those have been extracted from the complex dielectric function, that yields the refractive index of the respective material in a similar fashion. Fig. 15(e,f), shows that the above material has a homogeneous refractive index along the polarization direction $y$ & $z$, but it poses different over the $x$-direction, therefore this material showed up with two different indices 2.37 (along-$x$) and 2.31(along-$y, z$), hence it encoded itself with the optical birefringence. At limit $\omega \to 0$, the average refractive index of the material 2.34. The extinction coefficient K, from Fig. 15(f), vanishes below the band gap, nearly the band gap region it has a small peak and later it progresses around 5eV region and thereafter it decreases towards the 20eV with a single peak, and finally it diminishes. Maximum attenuation occurs at 5.16eV($x$-direction) and 4.96eV($y, z$-direction). Reflectivity was extracted from the refractive index and extinction coefficients, represented in Fig. 16, at static limit $\omega \to 0$, there was a 16% reflectivity (in case of $x$-direction polarization) and about 15% reflectivity for the other cases. Maximum reflectivity was achieved at 5.14eV for the $x$-orientation, which was about 32%, whether the other two cases had maximum reflectivity in the same energy level which of amount 31%.

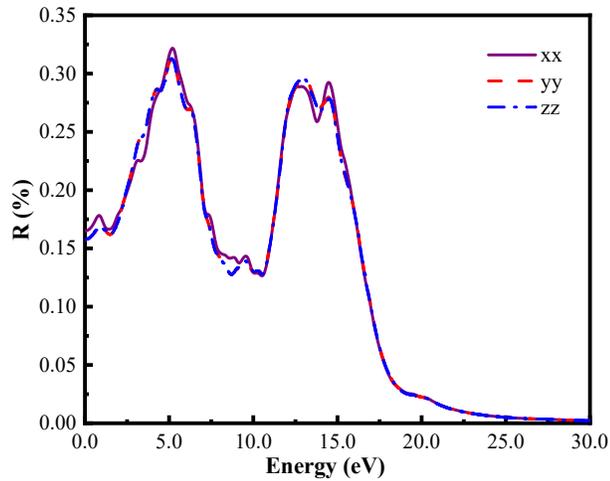

**Fig. 16**: Reflectivity of $Ni_{0.7}Cu_{0.3}Fe_{1.94}Al_{0.06}O_4$ nanoparticle.

4. **Conclusion**

Using the sol-gel auto-combustion method, a series of nanocrystalline $Ni_{0.7}Cu_{0.3}Al_xFe_{2-x}O_4$ has been produced with varying $Al^{3+}$ concentrations (0.00≤x≤0.10, in the step of 0.02) and annealed at 700ºC. X-ray diffraction patterns confirm that all investigated nanoparticles possess the same cubic single-phase structure with no impurities. The Rietveld refined lattice parameters of the annealed nanoparticles fall in the range (0.833-0.835 nm). It shows steady decrease in lattice parameter size with the increment of $Al^{3+}$ except for x=0.02 and 0.08. The average crystallite size of the investigated nanomaterials has been measured using Scherrer's formula considering the most prevalent XRD peaks (3 1 1), and the values are found in the range (61–71 nm) after the refinement. The VSM technique is employed to carry out the magnetic parameters. The Ni-Cu ferrite nanoparticles under this study are shown to be soft ferrimagnetic against an applied magnetic field. A decreasing trend is observed for shifts in saturation magnetization and Bohr magneton with $Al^{3+}$ incorporation. The saturation magnetization is observed as maximum for the sample with $Al^{3+}$ content (x=0.00). Moreover, the remanence ratio (0.00-0.094), coercivity (0.00-93.57 Oe), and the magnetic anisotropy constant (0.00-3612.19 erg/Oe) are varied with $Al^{3+}$ content. The purported high crystallinity and soft magnetic nature of the studied $Al^{3+}$ substituted Ni-Cu ferrite nanoparticles synthesized via the sol-gel method stand them potential candidates for multi-functional device applications. Moreover, the direct bandgap and the studied optical properties carried out through the DFT study suggest the investigated materials for optoelectronic device application, especially for their light-matter anisotropy response.

## Acknowledgment

The authors are thankful to the center of excellence of the Department of Mathematics and Physics at North South University (NSU), Dhaka 1229, Bangladesh. The NSU is supported this research through the NSU-OR research grant CTRG-20/SEPS/13.